\documentstyle[aps,preprint,tighten]{revtex}

\newcommand{\sn}{{\rm sn}}
\newcommand{\cn}{{\rm cn}}
\newcommand{\dn}{{\rm dn}}

\newcommand{\nc}{{\rm nc}}

\newcommand{\uT}{u_{\Theta}}
\newcommand{\ut}{u_{\theta}}
\newcommand{\vT}{\varphi_{\Theta}}

\draft

\begin{document}

\title{\bf Semiclassical Series at Finite Temperature}

\author{C. A. A. de Carvalho$^{1}$\footnote{e-mail: aragao@if.ufrj.br}
, R. M. Cavalcanti$^{2}$\footnote{e-mail: rmc@itp.ucsb.edu}, 
E. S. Fraga$^{1}$\footnote{e-mail: fraga@if.ufrj.br}, 
S. E. Jor\'as$^{1}$\footnote{e-mail: joras@if.ufrj.br}}

\address{$^1$ Instituto de F\'\i sica, Universidade Federal
do Rio de Janeiro, \\ C.P.~68528, Rio de Janeiro, RJ 21945-970, Brasil}

\address{$^2$ Institute for Theoretical Physics, University of 
California, \\ Santa Barbara, CA 93106-4030, USA}

\date{\today}

\maketitle

\begin{abstract}

We derive the semiclassical series for the partition function of a 
one-dimensional quantum-mechanical system consisting of a particle in 
a single-well potential. We do this by applying the method of steepest
descent to the path-integral representation of the partition function, 
and we present a systematic procedure to generate the terms of the
series using the minima of the Euclidean action as the only input.
For the particular case of a quartic anharmonic oscillator, we compute the 
first two terms of the series, and investigate their high and low 
temperature limits. We also exhibit the nonperturbative character of 
the terms, as each corresponds to sums over infinite subsets of 
perturbative graphs. We illustrate the power of such resummations by 
extracting from the first term an accurate nonperturbative estimate of the 
ground-state energy of the system and a curve for the specific heat. 
We conclude by pointing out possible 
extensions of our results which include field theories with spherically
symmetric classical solutions.

\end{abstract}


\section{Introduction}
\label{introduction}

Semiclassical series were introduced in Quantum Mechanics by the 
pioneering works of Brillouin \cite{brillouin}, 
Kramers \cite{kramers} and Wentzel \cite{wentzel}. In order to solve the 
time-independent Schr\"odinger equation for slowly-varying potentials, 
they proposed an ansatz for the wavefunction of the form
(here particularized to one dimension) 
\begin{equation}
\psi(x)=e^{iS(x)/\hbar},
\label{wave1}
\end{equation}
that led to a Ricatti equation 
for $S(x)$. The assumption of slow variation of the potential allowed 
for an iterative procedure in solving that equation. As a result, one 
obtained a series expansion of $S(x)$ in powers of $\hbar$, i.e., 
$S(x)=\sum_{n=0}^{\infty}S_n(x)\hbar^n$, with $S_{n+1}(x)$ given in 
terms of $S_n(x)$ by means of a recursion relation,
and $S_0(x)$ satisfying the 
Hamilton-Jacobi equation for a particle with the same energy and 
external potential as those of the Schr\"odinger equation. Thus, the 
zeroth order approximation could be associated to a classical path, from 
an arbitrary $x_0$ to $x$, whose classical action 
$S_0(x)=S_{\rm cl}(x_0\to x)$, apart from a sign, should be used in 
(\ref{wave1}). The choice of 
$x_0$ just fixed a normalization constant.

The terms of the series derived in references 
\cite{brillouin,kramers,wentzel} were plagued with divergences 
that occured at the turning points of the classical motion, a problem 
that came to be known as the connection problem. Dunham \cite{dunham} 
bypassed this difficulty by turning $x$ into a complex variable $z$, and 
using different linear combinations of $e^{\pm iS(x)/\hbar}$ as 
asymptotic series for $\psi(x)$ along the real axis. Assuming a potential 
with only two classical turning points $x_a$ and $x_b$ ($x_a < x_b$), 
three different linear combinations should be used as $x<x_a$, 
$x_a<x<x_b$ or $x>x_b$. Demanding that $\psi(z)$ be real, bounded and 
single-valued on the real axis, and writing $S(z)=\int^z p(z')\,dz'$, 
he could arrive at the quantization condition
\begin{equation}
\oint p(z) dz = 2\pi n\hbar\qquad(n=1,2,\ldots),
\label{quantization}
\end{equation}
for any closed contour encircling the segment $(x_a,x_b)$. The 
semiclassical series for $\psi(z)$ led to semiclassical series for 
$S(z)$ and $p(z)$ which, together with (\ref{quantization}), provided an 
expression for the energy levels as a power series in $\hbar$, 
generalizing the Bohr-Sommerfeld condition. That series was used as 
a starting point for the works of Bender, Olaussen and Wang 
\cite{olaussen} and Balian, Parisi and Voros \cite{parisi1,parisi2}, who 
refined such asymptotic representations to obtain highly accurate 
estimates of the energy levels of the single-well quartic anharmonic 
oscillator. These could be compared to the estimates obtained by Bender 
and Wu \cite{bender1,bender2,bender3} from the investigation of 
the large order behavior of perturbation theory 
\cite{lipatov,brezin,itzykson,zinnjustin}.

Apart from energy eigenfunctions and eigenvalues, semiclassical series 
were also written down for Green's functions by 
Balian and Bloch \cite{balian1}. Defining the resolvent operator
$\hat G$ through
\begin{equation}
(\hat H -z)\,\hat G = \openone
\label{hbalian}
\end{equation}
for a complex energy $z$, $\hat H$ being the Hamiltonian of the system, 
they used an ansatz of the form
\begin{equation}
G(x,x';z)=A(x,x')\,e^{S(x,x';z)/\hbar}.
\label{gbalian}
\end{equation}
{}From the defining equation (\ref{hbalian}), under assumptions 
similar to those in the first paragraph, they started from zeroth order 
estimates $S_0(x,x';z)$ and $A_0(x,x')$ by neglecting nonleading terms in 
$\hbar$. They found that $S_0$ satisfied a Hamilton-Jacobi equation for 
complex energy $z$, solved by a classical trajectory from $x$ 
to $x'$. The amplitude $A_0(x,x')$ turned out to be related to derivatives 
of the classical $S_0$. Just as before, the zeroth estimates could be used 
to generate a whole series expansion by an iterative procedure. Note 
that $z=E+i\gamma$, the analytic continuation to complex energies being 
used to circumvent problems with singularities (turning points in one 
dimension; caustics in higher dimensions).

In all the discussions mentioned, the first term of the 
semiclassical series was obtained from quantities related to some 
classical trajectory. Path integral representations for 
correlation functions \cite{feynman,schulman,rivers,weiss,kleinert}, which  
sum over trajectories, should therefore provide a natural setting to derive 
semiclassical series. Indeed, approximating quantum-mechanical path 
integrals by the stationary phase method leads to zeroth order estimates 
given by the paths that solve the classical equations of 
motion, the saddle points of the action functional. Thus,
\begin{equation}
\langle x | e^{-i\frac{\hat H}{\hbar}(t-t')}|x'\rangle=
\int_{x(t')=x'}^{x(t)=x}~[Dx(t)]~e^{iS[x(t)]/\hbar},
\label{bracket1}
\end{equation}
may be time Fourier transformed to yield (\ref{gbalian}). The 
saddle-point contribution, and fluctuations around it, will yield the 
zeroth order estimates $S_0$ and $A_0$ mentioned before.

Although semiclassical quantization by means of path integrals became 
widely used, as evidenced by Gutzwiller's work \cite{gutzwiller1}, and by 
extensions to field theories by Dashen, Hasslacher and Neveu
\cite{dashen1,dashen2,dashen3,rajaraman}, almost all discussions 
never went beyond the first term of a semiclassical series. Notable 
exceptions were the works of DeWitt-Morette 
\cite{morette1}, for arbitrary potentials 
in Quantum Mechanics, and Mizrahi \cite{mizrahi}, for the single-well 
quartic anharmonic oscillator in Quantum Mechanics (more recently
\cite{roncadelli}, a derivation of the higher order terms of the series 
using transport-type recurrence equations also became available).
However, these 
contributions to the mathematical physics literature did not receive 
the attention they certainly deserved in more applied work. We were no 
exception, and only became aware of these articles after we had rederived 
the semiclassical series using the methods of section 
\ref{statistical}.

Semiclassical methods for finite temperature field theories 
\cite{bernard,dolan,weinberg} also remained restricted to derivations of 
the first term of a semiclassical series \cite{ma}, even when the 
problem was reduced to Quantum Statistical Mechanics 
\cite{harrington,dolankiskis}, 
viewed as field theory at a point (zero spatial dimension). Some 
references resorted to extensions to the complex plane 
\cite{mottola,carlitz,wileyboyanholman} 
to include complex paths required to describe Fourier transformed 
quantities but, again, those treatments were not concerned with 
obtaining the whole series.

The central objective of the present article is to bridge this gap. 
Thus, we undertook the task of finding a systematic path integral 
procedure to generate a 
semiclassical series for Quantum Statistical Mechanics which led us to 
the construction of each term of the series from the knowledge of the 
solution(s) of 
the classical equations of motion. We concentrated our attention on the 
partition function, for which use of the method of steepest descent only 
required {\it real} solutions as saddle-points \cite{joras}. The 
restriction of our analysis to one-dimensional quantum-mechanical 
systems (i.e., scalar field theories at a point and at finite temperature) 
allowed us to construct the semiclassical propagator needed to generate the 
terms of the series.

This article is organized as follows: section \ref{statistical} 
presents the derivation of the 
semiclassical series for a generic potential of the single-well type, 
both in quantum-mechanical language and in field-theoretic language, 
the latter allowing for a simple connection with the works of references 
\cite{morette1,mizrahi}. It should be 
remarked, however, that our presentation is quite simple, being a natural 
extension of textbook material \cite{zinnjustin}, and having profitted 
greatly from 
the clear account of reference \cite{wileyboyanholman}. To emphasize that our 
construction is free of turning-point singularities, details of the 
derivation were worked out in Appendices \ref{A} and \ref{B}; 
section \ref{quartic} discusses the single-well quartic anharmonic 
oscillator. There, we constructed the two first terms of the series 
explicitly, and looked at relevant limits. We also computed the ground-state 
energy and the specific heat, as illustrations. Appendices \ref{C}, \ref{D} 
and \ref{E} complement the calculations in the text; section \ref{conclusions} 
presents our conclusions, points 
out directions for future work, and lists a number of situations where 
our results could be applied.

\section{Statistical Mechanics}     
\label{statistical}

\subsection{Quantum-mechanical path integrals}    
\label{qmpath}

The partition function for a one-dimensional quantum-mechanical 
system consisting of a particle of mass $m$ in the presence of a 
potential $V(x)$ in equilibrium at inverse temperature $\beta$ can 
be written as a path integral:
\begin{equation}
Z(\beta)=\int_{-\infty}^\infty dx_0 \int_{x(0)=x_0}^{x(\beta\hbar)=x_0}
[{\cal D}x(\tau)]\, e^{-S/\hbar},
\end{equation}
\begin{equation}
S[x]=\int_0^{\beta\hbar} d\tau
\left[\frac{1}{2}\,m\left(\frac{dx}{d\tau}\right)^2+V(x)\right].
\end{equation}
For convenience we define the dimensionless quantities 
$q\equiv x/x_N$, $\theta\equiv \omega_N\tau$, 
$\Theta\equiv \beta\hbar\omega_N$, $U(q)\equiv V(x_Nq)/m\omega_N^2x_N^2$
and $g\equiv\hbar/m\omega_Nx_N^2$,
where $\omega_N^{-1}$ and $x_N$
are natural time and length scales of the problem, respectively.
In terms of these quantities we rewrite the partition function as
\begin{equation}
Z(\Theta)=\int_{-\infty}^{\infty}dq_0
\int_{q(0)=q_0}^{q(\Theta)=q_0}[{\cal D}q(\theta)]\, e^{-I/g},
\label{Z}
\end{equation}
\begin{equation}
I[q]=\int_0^\Theta d\theta \left[\frac{1}{2}\,{\dot q}^2
+U(q)\right],
\end{equation}
where the dot denotes differentiation in $\theta$. 

We generate a semiclassical series for $Z(\Theta)$ by: (i) finding the minima 
$q_c(\theta)$ of the Euclidean action $I$, i.e., the stable classical 
paths that solve the Euler-Lagrange equation of motion, subject to the 
boundary conditions; (ii) expanding the Euclidean action around these classical 
paths; (iii) deriving a quadratic semiclassical propagator by neglecting 
terms higher than second order in the expansion; (iv) using that propagator 
to compute higher (than quadratic) order contributions perturbatively.

For the sake of simplicity, we shall restrict our analysis to potentials 
of the single-well type, twice differentiable, and such that $U'(q)=0$
only at the minimum of $U$, which we shall assume to be at the origin 
(see Fig.\ \ref{potential}). This guarantees that, 
given $q_0$ and $\Theta$, there will be a {\em unique} classical path 
satisfying the boundary conditions. Multiple-well potentials force us 
to consider more than one classical path for certain choices of $q_0$ 
and $\Theta$. This phenomenon has been analyzed, for a double-well type 
potential, using the language of catastrophes and bifurcations \cite{bjp}. 
Semiclassical series for the double-well quartic oscillator will be 
presented elsewhere \cite{joras}.

The Euler-Lagrange equation ($U'\equiv dU/dq$)
\begin{equation}
\ddot{q} -U'(q)=0,
\label{euler}
\end{equation}
subject to the boundary conditions $q(0)=q(\Theta)=q_0$, describes the
motion of a particle in the potential {\em minus} $U$. Its first integral is
\begin{equation}
\frac{1}{2}\,{\dot q}^2=U(q)-U(q_t),
\label{motion}
\end{equation}
where $q_t$ denotes the single turning point (since we have an inverted single 
well) of the motion, defined implicitly by
\begin{equation}
\Theta=2\int_{q_0}^{q_t} \frac{dq}{v(q,q_t)},
\label{theta}
\end{equation}
where $v(q,q')\equiv {\rm sign}(q'-q)\sqrt{2[U(q)-U(q')]}$, and equation 
(\ref{theta}) is a consequence of integrating (\ref{motion}). Thus, for  
a single well, given $q_0$ and $\Theta$, the classical path will go from 
$q_0$, at $\theta=0$, to $q_t=q_t(q_0,\Theta)$, at $\theta=\Theta/2$, and 
return to $q_0$ at $\theta=\Theta$. (Note that 
${\rm sign}(q_t)={\rm sign}(q_0)$.)

The action for this classical path has a simple expression in terms of
its turning point:
\begin{equation}
I[q_c]=\Theta\, U(q_t)+2\int_{q_0}^{q_t}dq \, v(q,q_t),
\label{action}
\end{equation}
where we have used (\ref{motion}). The first term in (\ref{action}) 
corresponds to the high-temperature limit of $Z(\Theta)$, where classical 
paths collapse to a point ($q_t\to q_0$). The last term will be 
negligible for potentials that vary little over a thermal wavelength 
$\lambda=\hbar\sqrt{\beta/m}$. However, by decreasing the temperature 
it will become important and bring in quantum effects.

We now expand the action around the classical path. Letting 
$q(\theta)=q_c(\theta)+\eta(\theta)$, with 
$\eta(0)=\eta(\Theta)=0$, we obtain
\begin{equation}
I[q]=I[q_c]+I_2[\eta]+\delta I[\eta],
\label{fluc}
\end{equation}
where
\begin{equation}
I_2[\eta] \equiv \frac{1}{2}\int_0^\Theta d\theta\,\left\{
\dot\eta^2(\theta)+U''[q_c(\theta)]\,\eta^2(\theta)\right\},
\end{equation}
\begin{equation}
\delta I[\eta] \equiv \int_0^\Theta d\theta\, \delta U(\theta,\eta)
=\sum_{n=3}^\infty \frac{1}{n!}\int_0^\Theta d\theta\,
U^{(n)}[q_c(\theta)]\,\eta^n(\theta).
\label{deltaI}
\end{equation}
Inserting (\ref{fluc}) into (\ref{Z}) and expanding $e^{-\delta I/g}$
in a power series yields
\begin{equation}
Z(\Theta)=\int_{-\infty}^\infty dq_0\,e^{-I[q_c]/g}
\int_{\eta(0)=0}^{\eta(\Theta)=0}
[{\cal D}\eta(\theta)]\, e^{-I_2[\eta]/g}
\sum_{m=0}^\infty \frac{1}{m!}\left(-\frac{\delta I[\eta]}{g}\right)^m.
\label{zpow}
\end{equation}
If we interchange, 
\`a la Feynman \cite{feynman}, the integral over deviations from the
classical path with the 
integrals over various times $\{\theta_i\}$, coming from insertions of 
(\ref{deltaI}), we break up the path-integral into pieces which go from 
zero to the various $\{\theta_i\}$ and, finally, to $\Theta$. Thus,
\begin{equation}
Z(\Theta)=\int_{-\infty}^\infty dq_0\, e^{-I[q_c]/g}\left[
G_c(0,0;\Theta,0)+
\sum_{n=1}^\infty \frac{(-1)^n}{g^n n!}\,G_n(q_0,\Theta)\right],
\label{zbeta}
\end{equation}
\begin{eqnarray}
G_n(q_0,\Theta)&=&\left(\prod_{j=1}^n\,\int_0^\Theta d\theta_j 
\int_{-\infty}^\infty d\eta_j \right)
G_c(0,0;\theta_1,\eta_1)\,\delta U(\theta_1,\eta_1)
\nonumber \\
& &\times G_c(\theta_1,\eta_1;\theta_2,\eta_2)\cdots\delta U(\theta_n,\eta_n)
\,G_c(\theta_n,\eta_n;\Theta,0)
\Big|_{\theta_1\le\theta_2\le\ldots\le\theta_n},
\end{eqnarray}
with $G_c(\theta_1,\eta_1;\theta_2,\eta_2)$, the semiclassical propagator,
emerging naturally from doing the path-integral in each piece:
\begin{equation}
G_c(\theta_1,\eta_1;\theta_2,\eta_2)=
\int_{\eta(\theta_1)=\eta_1}^{\eta(\theta_2)=\eta_2} 
[{\cal D}\eta(\theta)]\, e^{-I_2[\theta_1,\theta_2;\eta]/g},
\label{gc}
\end{equation}
\begin{equation}
I_2[\theta_1,\theta_2;\eta]=\frac{1}{2}\int_{\theta_1}^{\theta_2} d\theta\, 
\left\{{\dot \eta}^2 +U''[q_c(\theta)]\, \eta^2\right\}.
\label{i2}
\end{equation}

It remains to show how one can obtain that propagator from the knowledge
of the classical path. This is the central point in the whole procedure. 
For this, we use the fact that the action $I_2$ is quadratic
in $\eta$, and so the path integral in (\ref{gc})
is completely determined by the 
extremum $\eta_e(\theta)$ of $I_2[\theta_1,\theta_2;\eta]$, which satisfies
\begin{equation}
\ddot \eta - U''[q_c(\theta)]\,\eta=0,
\label{extremum}
\end{equation}
subject to the boundary conditions $\eta(\theta_1)=\eta_1$ and 
$\eta(\theta_2)=\eta_2$. Thus,
\begin{equation}
G_c(\theta_1,\eta_1;\theta_2,\eta_2)=G_c(\theta_1,0;\theta_2,0)\,
e^{-I_2[\theta_1,\theta_2;\eta_e]/g},
\label{gc2}
\end{equation}
where, after an integration by parts, 
\begin{equation}
I_2[\theta_1,\theta_2;\eta_e]=\frac{1}{2}\left[\eta_2\,\dot\eta_e(\theta_2)
-\eta_1\,\dot\eta_e(\theta_1)\right].
\end{equation}

We can obtain $\eta_e(\theta)$ by finding the linear combination of any two 
linearly independent solutions, $\eta_a(\theta)$ and $\eta_b(\theta)$, of 
(\ref{extremum}) which satisfies $\eta_e(\theta_1)=\eta_1$ and 
$\eta_e(\theta_2)=\eta_2$. The result is
\begin{equation}
\eta_e(\theta)=\frac{\eta_1\,\Omega(\theta,\theta_2)
+\eta_2\,\Omega(\theta_1,\theta)}{\Omega(\theta_1,\theta_2)},
\end{equation}
where
\begin{equation}
\Omega(\theta,\theta')\equiv \eta_a(\theta)\,\eta_b(\theta')
-\eta_a(\theta')\,\eta_b(\theta).
\label{omega}
\end{equation}
We may then write
\begin{equation}
I_2[\theta_1,\theta_2;\eta_e]=\frac{1}{2\,\Omega_{12}}\,[W_{12}\,\eta_2^2
+W_{21}\,\eta_1^2-(W_{11}+W_{22})\,\eta_1\eta_2],
\label{i2'}
\end{equation}
where $\Omega_{ij}\equiv\Omega(\theta_i,\theta_j)$ and
$W_{ij}\equiv\partial\Omega_{ij}/\partial\theta_j$.
(Note that $W_{ii}$ is the Wronskian of $\eta_a$ and 
$\eta_b$ computed at $\theta_i$.)

Explicit expressions for $\eta_a(\theta)$ and $\eta_b(\theta)$
can be obtained as follows.
By differentiating (\ref{euler}) with respect to $\theta$, one can verify 
that $\eta_a(\theta)=\dot q_c(\theta)$ satisfies (\ref{extremum}).
For the second solution, we take
$\eta_b(\theta)=\dot q_c(\theta)\,Q(\theta)$,
where $Q(\theta)$ is defined as
\begin{equation}
Q(\theta)=Q(0)+\int_0^\theta\frac{d\theta'}{{\dot q}^2_c(\theta')}
\label{qthetaq0}
\end{equation}
for $\theta<\Theta/2$, $Q(\theta)=-Q(\Theta-\theta)$ for
$\theta>\Theta/2$, and $Q(0)$ is chosen so as to make
$\dot\eta_b(\theta)$ continuous at $\theta=\Theta/2$ (see Appendix A). 
One can easily check, using (\ref{euler}), that $\eta_b(\theta)$ 
indeed satisfies (\ref{extremum}). (Alternatively, one could use a procedure 
introduced by Cauchy \cite{morette1,mizrahi}, and differentiate the classical 
solution $q_c(\theta)$ with respect to any two parameters related to its two 
constants of integration.) We can now write explicit expressions
for $\Omega_{12}$ and $W_{ij}$:
\begin{equation}
\Omega_{12}=\dot q_c(\theta_1)\,\dot q_c(\theta_2)\,
[Q(\theta_2)-Q(\theta_1)],
\label{o12}
\end{equation}
\begin{equation}
W_{ij}=\dot q_c(\theta_i)\, U'[q_c(\theta_j)]\,[Q(\theta_j)-Q(\theta_i)]
+\frac{\dot q_c(\theta_i)}{\dot q_c(\theta_j)}.
\label{w12}
\end{equation}

As a final step, the pre-factor in (\ref{gc2}) is derived in 
Appendix \ref{B}, using the methods of 
Refs.\ \cite{zinnjustin,wileyboyanholman}. The result is 
\begin{equation}
G_c(\theta_1,0;\theta_2,0)={\left[\frac{W_{11}}{2\pi g\,\Omega_{12}}
\right]}^{1/2}.
\end{equation}
{}From (\ref{w12}), one easily finds $W_{ii}=1$. Therefore, our quadratic 
semiclassical propagator is given by 
\begin{equation}
G_c(\theta_1,\eta_1;\theta_2,\eta_2)=\frac{1}{\sqrt{2\pi g\,\Omega_{12}}}\,
\exp\left[-\frac{1}{2g\,\Omega_{12}}\,(W_{12}\,\eta_2^2
+W_{21}\,\eta_1^2-2\,\eta_1\eta_2)\right].
\label{gs}
\end{equation}
As promised, it is completely determined by the classical solution. 

Finally, we note that the van Vleck determinant $\Delta$ is a 
by-product of (\ref{gs}):
\begin{equation}
\Delta(q_0,\Theta)=G_c^{-2}(0,0;\Theta,0)=2\pi g\,\Omega(0,\Theta)
=4\pi g\,\dot{q}_c^2(0)\,Q(0).
\label{vanvleck}
\end{equation}
Using (\ref{motion}) and (\ref{id2}) one can express $\Delta$ as
\begin{equation}
\Delta=\frac{4\pi g\,[U(q_t)-U(q_0)]}{U'(q_t)}\left(\frac{\partial\Theta}
{\partial q_t}\right)_{q_0}.
\label{Delta2}
\end{equation}
Together with (\ref{action}), this shows that one does not need to
know $q_c(\theta)$ in order to write the first 
term in the semiclassical series (\ref{zbeta}); it is enough to know
$q_t(q_0,\Theta)$.


\subsection{Field-theoretic approach}
\label{ftpath}

We may generate the semiclassical series in an alternative way, 
which connects it quite naturally to the diagrammatics of field theory.

The summation in (\ref{zpow}) can be written more explicitly as
\begin{equation}
\sum_{m=0}^\infty \frac{1}{m!}\left(-\frac{\delta I[\eta]}{g}\right)^m=1+
\sum_{m=1}^\infty\frac{(-1)^m}{g^m m!}\prod_{j=1}^m\left[\sum_{n_j=3}^\infty
\frac{1}{n_j!}\int_0^\Theta d\theta_j\, U^{(n_j)}[q_c(\theta_j)]
\,\eta^{n_j}(\theta_j)\right].
\label{sumexp}
\end{equation}
As a consequence, one is led to compute integrals of the following type:
\begin{equation}
\langle\eta(\theta_1)\cdots\eta(\theta_k)\rangle\equiv
\int_{\eta(0)=0}^{\eta(\Theta)=0}[{\cal D}\eta(\theta)] 
\,e^{-I_2[0,\Theta;\eta]/g}\,\eta(\theta_1)\cdots\eta(\theta_k).
\label{etaeta}
\end{equation}
Such integrals emerge naturally as functional 
derivatives of the following generating functional:
\begin{equation}
{\cal Z}[J]=
\int_{\eta(0)=0}^{\eta(\Theta)=0}[{\cal D}\eta(\theta)] 
\,e^{-\frac{1}{g}\left\{I_2[0,\Theta;\eta]-\int_{0}^{\Theta}
d\theta\, J(\theta)\,\eta(\theta)\right\}}.
\label{gtil}
\end{equation}
Indeed,
\begin{equation}
\langle\eta(\theta_1)\cdots\eta(\theta_k)\rangle =
g^k\,\frac{\delta^k\,{\cal Z}[J]}
{\delta J(\theta_1)\cdots\delta J(\theta_k)}\Bigg|_{J=0}.
\label{bracket}
\end{equation}

In order to compute ${\cal Z}[J]$, we define
\begin{equation}
\eta(\theta)=\tilde\eta(\theta)+\int_{0}^{\Theta}d\theta'\, 
{\cal G}(\theta,\theta')\,J(\theta'),
\label{tileta}
\end{equation}
where $\tilde\eta(0)=\tilde\eta(\Theta)=0$, and ${\cal G}(\theta,\theta')$
satisfies
\begin{equation}
\left\{-\frac{\partial^2}{\partial\theta^2}+U''[q_c(\theta)]\right\}
{\cal G}(\theta,\theta')=\delta(\theta-\theta'),
\qquad{\cal G}(0,\theta')={\cal G}(\Theta,\theta')=0.
\label{green}
\end{equation}
Inserting (\ref{tileta}) in (\ref{gtil}), and noting that
$[{\cal D}\eta(\theta)]=[{\cal D}\tilde\eta(\theta)]$,
we obtain
\begin{equation}
{\cal Z}[J]=e^{\frac{1}{2g}\int_0^{\Theta}d\theta
\int_0^{\Theta}d\theta'\,J(\theta)\,{\cal G}(\theta,\theta')\,
J(\theta')}\int_{\tilde\eta(0)=0}^{\tilde\eta(\Theta)=0}
[{\cal D}\tilde\eta(\theta)]\,e^{-I_2[0,\Theta;\tilde\eta]/g}
\label{ZJ}
\end{equation}
The path integral in (\ref{ZJ}) has the same form as the one
in (\ref{gc}), so we finally arrive at
\begin{equation}
{\cal Z}[J]=G_c(0,0;\Theta,0)
\,\exp\left[\frac{1}{2g}\int_{0}^{\Theta}d\theta \int_{0}^{\Theta}d\theta' 
J(\theta)\,{\cal G}(\theta,\theta')\, J(\theta')\right].
\end{equation}
Using this result, we can now calculate (\ref{bracket}).
The result is simply
\begin{equation}
\langle\eta(\theta_1)\cdots\eta(\theta_k)\rangle =g^{k/2}\,
G_c(0,0;\Theta,0)\,
\sum_P{\cal G}(\theta_{i_1},\theta_{i_2})\cdots 
{\cal G}(\theta_{i_{k-1}},\theta_{i_k}),
\label{sumP}
\end{equation}
if $k$ is even, and zero otherwise. $\sum_P$ denotes sum over all 
possible pairings of the $\theta_{i_j}$. Inserting this into (\ref{zpow}) 
and (\ref{sumexp}) yields the semiclassical series for $Z(\Theta)$.

We still have to solve Eq.\ (\ref{green}). This can be easily
done if one notes that, for
$\theta\ne\theta'$, it has the same form
as Eq.\ (\ref{extremum}). Therefore, ${\cal G}(\theta,\theta')$
can be constructed, just as $\eta_e(\theta)$ itself, as a linear 
combination of the $\eta_a(\theta)$ and $\eta_b(\theta)$ defined previously:
\begin{equation}
{\cal G}(\theta,\theta')=\left\{
\begin{array}{ll}
a_-\eta_a(\theta)+b_-\eta_b(\theta),& \theta<\theta' \\
a_+\eta_a(\theta)+b_+\eta_b(\theta),& \theta>\theta'.
\end{array}
\right.
\label{calG}
\end{equation}
Continuity imposes
\begin{equation}
{\cal G}(\theta'+\epsilon,\theta')={\cal G}(\theta'-\epsilon,\theta'),
\label{cont}
\end{equation}
whereas (\ref{green}) leads to
\begin{equation}
\frac{\partial}{\partial\theta}\,{\cal G}(\theta,\theta')
\Big|_{\theta=\theta'+\epsilon} 
- \frac{\partial}{\partial\theta}\,{\cal G}(\theta,\theta')
\Big|_{\theta=\theta'-\epsilon}=-1,
\label{deriv}
\end{equation}
with $\epsilon\to 0^+$. (\ref{cont}), (\ref{deriv}) and the
boundary conditions completely determine the coefficients
in (\ref{calG}). The final result is
\begin{equation}
{\cal G}(\theta,\theta')=\frac{\Omega(0,\theta_<)\,
\Omega(\theta_>,\Theta)}{\Omega(0,\Theta)},
\label{wronsk}
\end{equation}
where $\theta_<(\theta_>)\equiv{\rm min(max)}\{\theta,\theta'\}$,
and $\Omega(\theta_1,\theta_2)$ is the function defined in (\ref{omega}).

\section{The Single-well Quartic Oscillator}
\label{quartic}

In this section, we study the potential 
\begin{equation}
V(x)=\frac{1}{2}\,m\omega^2x^2+\frac{1}{4}\,\lambda x^4.
\label{vx}
\end{equation}
Choosing $\omega_N=\omega$ and $x_N=\sqrt{m\omega^2/\lambda}$, and
introducing the dimensionless quantities 
of section \ref{statistical}, we have
$g=\lambda\hbar/m^2\omega^3$ and
\begin{equation}
U(q)=\frac{1}{2}\,q^2+\frac{1}{4}\,q^4.
\label{ux}
\end{equation}

Integrating (\ref{motion}) leads to \cite{grads,byrd}
\begin{equation}
q_c(\theta)=q_t\, \nc (u_{\theta},k),
\label{classical}
\end{equation}
where $\nc (u,k)\equiv 1/\cn  (u,k)$ is one of the Jacobian Elliptic 
functions \cite{grads,byrd,as}, and
\begin{equation}
u_{\theta}=\sqrt{1+q_t^2}\left(\theta-\frac{\Theta}{2}\right),
\qquad k=\sqrt{\frac{2+q_t^2}{2\,(1+q_t^2)}}.
\label{u,k}
\end{equation}
For future use, we note that (\ref{u,k}) can be rewritten as
\begin{equation}
u_{\theta}=\frac{2\theta-\Theta}{2\sqrt{2k^2-1}},
\qquad|q_t|=\sqrt{\frac{2\,(1-k^2)}{2k^2-1}}.
\label{gqt}
\end{equation}
The relation between $q_0$ and $q_t$ is obtained by taking
$\theta=\Theta$ in (\ref{classical}):
\begin{equation}
q_0=q_c(\Theta)=q_t\,\nc\,\uT.
\label{q0-qt}
\end{equation}
(We shall often omit the $k$-dependence in the Jacobian Elliptic
functions.)

The action for the classical path (\ref{classical}) is
\begin{equation}
I[q_c]=\Theta\, U(q_t)+\sqrt{2}\int_{|q_t|}^{|q_0|}dq\,
\sqrt{(q^2+q_t^2+2)(q^2-q_t^2)}.
\end{equation}
Performing the integral [Ref.\ \cite{grads}, formula 3.155.6]
and replacing $q_0$ by the r.h.s.\ of (\ref{q0-qt}), we obtain
\begin{eqnarray}
I[q_c]&=&\Theta\left(\frac{1}{2}\,q_t^2+
\frac{1}{4}\,q_t^4\right)+\frac{4}{3}\left\{-\sqrt{1+q_t^2}
\left[{\rm E}(\vT,k)+\frac{1}{2}\,q_t^2\,\uT\right]\right.
\nonumber \\
& &+\left.\sn\,\uT\left(1+\frac{1}{2}\,q_t^2\,\nc^2\uT
\right)\sqrt{1+\frac{1}{2}\,q_t^2\,(1+\nc^2\uT)}\right\},
\label{iqc}
\end{eqnarray}
where ${\rm E}(\varphi,k)$ denotes the Elliptic Integral of the 
Second Kind and
$\varphi_{\theta}\equiv\arccos[q_c(\theta)/q_0]=\arccos(\cn\,\ut)$.

For the construction of the quadratic semiclassical propagator we shall need
\begin{equation}
\eta_a(\theta)=\dot q_c(\theta)=q_t\,\sqrt{1+q_t^2}\,
\sn\, u_{\theta}\,\dn\, u_{\theta}\,\nc ^2u_{\theta}
\label{qatheta}
\end{equation}
and
\begin{eqnarray}
Q(\theta)&=&q_t^{-2}(1+q_t^2)^{-3/2}\left[\left(1-\frac{1}{k^2}\right)
u_{\theta}+\left(\frac{1}{k^2}-2\right){\rm E}(\varphi_{\theta},k)\right.
\nonumber \\
& &-\left.\frac{\cn\,u_{\theta}\,\dn\,u_{\theta}}{\sn\,u_{\theta}}
+(k^2-1)\,\frac{\cn\,u_{\theta}\,\sn\,u_{\theta}}{\dn\,u_{\theta}}\right].
\label{qtheta}
\end{eqnarray}
We may then obtain $\eta_b(\theta)=\dot q_c(\theta)\,Q(\theta)$ and, thus, 
$\Omega_{12}$ and $W_{12}$ from (\ref{o12}) and (\ref{w12}). Finally, 
use of (\ref{gs}) will yield the desired propagator .

For the series expansion of the partition function (\ref{zbeta}), 
we shall need
\begin{equation}
\delta U(\theta,\eta)=q_c(\theta)\,\eta^3+\frac{1}{4}\,\eta^4,
\label{deltaU}
\end{equation}
obtained from (\ref{deltaI}). Therefore, we have to consider not only 
the usual quartic vertex, but an additional time($\theta$)-dependent 
cubic term. This completes the set of ingredients needed to write down 
a semiclassical series for any correlation. In the next subsection, we 
shall concentrate on the first term of the series (\ref{zbeta}) for 
$Z(\Theta)$, which yields the quadratic approximation.


\subsection{The quadratic approximation for $Z(\Theta)$}
\label{quadratic}
	
{}From the knowledge of the classical action and of the Van Vleck 
determinant, we define
\begin{equation}
Z_2(\Theta)\equiv \int_{-\infty}^\infty dq_0\, e^{-I[q_c]/g} 
\Delta^{-1/2}
\label{z2.1}
\end{equation}
as the quadratic approximation to $Z(\Theta)$. To perform the
integral over $q_0$ one must write $I[q_c]$ and $\Delta$ solely
in terms of $q_0$ (and $\Theta$), but except in rare cases
this is not an easy task. Usually, it is much simpler to write
these quantities in terms of $q_t$ [see Eq.\ (\ref{q0-qt}) and
Appendix \ref{C}],
and so it is natural to trade $q_0$ for $q_t$ as the 
integration variable in (\ref{z2.1}). This is much simplified
by the fact that the Jacobian of the map $q_0\to q_t$ is
simply related to the van Vleck determinant. In fact, Eqs.\ (\ref{theta})
and (\ref{Delta2}) imply
\begin{equation}
\left(\frac{\partial q_0}{\partial q_t}\right)_\Theta 
=-\frac{(\partial\Theta/\partial q_t)_{q_0}}
{(\partial\Theta/\partial q_0)_{q_t}}
=\frac{1}{2}\,v(q_0,q_t) 
\left(\frac{\partial \Theta}{\partial q_t}\right)_{q_0}
=-\frac{U'(q_t)\,\Delta}{4\pi g\, v(q_0,q_t)}.
\end{equation}
Eq.\ (\ref{z2.1}) then becomes
\begin{equation}
Z_2(\Theta)= -\frac{1}{4\pi g}\int_{q_{\Theta}^{-}}^{q_{\Theta}^{+}} 
dq_t\,\frac{U'(q_t)\,\Delta^{1/2}}{v(q_0,q_t)}\, e^{-I[q_c]/g}\equiv
\int_{q_{\Theta}^{-}}^{q_{\Theta}^{+}} dq_t\,D(q_t,\Theta)\,e^{-I[q_c]/g},
\label{z2.2}
\end{equation}
where $q_{\Theta}^{\pm}\equiv\lim_{q_0\to\pm\infty}q_t(q_0,\Theta)$. 

The expression above is valid for single-well potentials in general.
Now, let us especialize to the potential (\ref{ux}). $I[q_c]$ is
given by (\ref{iqc}), and using (\ref{vanvleck}) and (\ref{qtheta})
one can write $D(q_t,\Theta)$ as
\begin{eqnarray}
D(q_t,\Theta)&=&\frac{(1+q_t^2)^{1/4}}{\sqrt{4\pi g}}
\left[\frac{1-k^2}{k^2}\,\uT
+\frac{2k^2-1}{k^2}\,E(\vT,k)\right.
\nonumber \\
& &+\left.\frac{\cn\,\uT\,
\dn\,\uT}{\sn\,\uT}+(1-k^2)\,\frac{\cn\,\uT\,
\sn\,\uT}{\dn\,\uT}\right]^{1/2}.
\label{Dqt}
\end{eqnarray}
{}From (\ref{q0-qt}) it follows that $q_0\to\infty$ when 
$\cn(\uT,k)=0$, which occurs when $\uT={\rm K}(k)$, where
${\rm K}(k)$ is the Complete Elliptic Integral of the First Kind.
Using (\ref{gqt}), this condition can be written as an equation in $k$:
\begin{equation}
\frac{\Theta}{2\sqrt{2k^2-1}}={\rm K}(k).
\label{utheta}
\end{equation}
The graph of $f(k)\equiv 2\sqrt{2k^2-1}\,{\rm K}(k)$ is plotted in
Fig.\ \ref{ktheta}. It increases monotonically from zero (at $k=1/\sqrt{2}$)
to infinity (as $k\to 1$), and so for each nonnegative value of $\Theta$
Eq.\ (\ref{utheta}) has a unique solution, which we denote
by $k_{\Theta}$. 
Eq.\ (\ref{gqt}) then gives the corresponding value of $q_{\Theta}^{+}$
($q_{\Theta}^{-}=-q_{\Theta}^{+}$, since $U(-q)=U(q)$).


\subsection{Limiting cases of the quadratic approximation}

Expression (\ref{z2.2}) may be used to compute $Z_2(\Theta)$ numerically 
for any value of $\Theta$. However, certain limiting cases may be
dealt with analytically, and it is instructive to consider them first
in order to recover some known results, thus providing a consistency
check. These limits are: (1) the harmonic oscillator
($g\to 0$), (2) high temperatures ($\Theta\to 0$), and (3) low
temperatures ($\Theta\to\infty$).


\subsubsection{The harmonic oscillator}

Since $V(x)=\frac{1}{2}\,m\omega^2x^2$ when $g=0$, one should obtain
the partition function of the harmonic oscillator in the limit $g\to 0$.
To see how to arrive at this result starting from the expressions derived
in this section, we note that
in this limit one can perform the integral (\ref{z2.2}) using the
steepest descent method. For this, we only need the first nontrivial
term in the series expansion of $I[q_c]$ and 
$D(q_t,\Theta)$ around $q_t=0$.

When $|q_t|\ll 1$, one has $k\approx 1-\frac{1}{4}\,q_t^2$
and ${k'}^2\equiv 1-k^2\approx\frac{1}{2}\,q_t^2\ll 1$. Now, we use
the smallness of $k'$ to derive an approximation for 
${\rm E}(\vT,k)$. The first step is to write it in
terms of $k'$:
\begin{equation}
{\rm E}(\varphi,k)=\int_0^{\varphi}\sqrt{1-k^2\sin^2x}\,dx
=\int_0^{\varphi}\cos x\,\sqrt{1+{k'}^2\tan^2x}\,dx.
\end{equation}
Expanding the integrand to first order in ${k'}^2$, integrating
the result, and using the fact that $\sin\vT=\sn\,\uT$
and $\cos\vT=\cn\,\uT$, we finally obtain
\begin{equation}
{\rm E}(\vT,k)\approx\sn\,\uT+\frac{1}{4}\,q_t^2
\left[\ln\left(\frac{1+\sn\,\uT}{\cn\,\uT}\right)-\sn\,\uT\right].
\label{approxE}
\end{equation}
Inserting this result in (\ref{iqc}) and expanding the square-roots
in that equation in powers of $q_t$, one obtains
\begin{equation}
I[q_c]=q_t^2\left[\frac{1}{2}\,\Theta-\frac{2}{3}\,\uT
-\frac{1}{3}\,\ln\left(\frac{1+\sn\,\uT}{\cn\,\uT}\right)
+\sn\,\uT\,\nc^2\uT\right]+{\cal O}(q_t^4).
\end{equation}
Now, we use the fact that $\uT\to\Theta/2$ and $k\to 1$ when $q_t\to 0$,
and $\sn(u,1)=\tanh u$ and $\cn(u,1)={\rm sech}\,u$,
thus finally arriving at
\begin{equation}
I[q_c]=\frac{1}{2}\,q_t^2\,\sinh\Theta+{\cal O}(q_t^4).
\label{Ig}
\end{equation}

In the case of $D(q_t,\Theta)$ it is enough to take the limit $q_t\to 0$ in
(\ref{Dqt}). Then, the first and the last term in square brackets
vanish, the second approaches $\tanh(\Theta/2)$ [use (\ref{approxE})],
and the third approaches ${\rm sech}^2(\Theta/2)\,\coth(\Theta/2)$.
Combining these results, one obtains
\begin{equation}
D(q_t,\Theta)=[4\pi g\,\tanh(\Theta/2)]^{-1/2}+{\cal O}(q_t^2).
\label{Dg}
\end{equation}

Inserting (\ref{Ig}) and (\ref{Dg}) in (\ref{z2.2}) yields the desired
small-$g$ limit of the partition function:
\begin{equation}
Z_2(\Theta)\stackrel{g\to 0}{\sim}
\int_{-\infty}^{\infty}\frac{e^{-(1/2g)\,q_t^2\,\sinh\Theta}}
{\sqrt{4\pi g\,\tanh(\Theta/2)}}\,dq_t=\frac{1}{2\,\sinh(\Theta/2)}.
\label{z2lim}
\end{equation}
%


\subsubsection{High temperatures}

At high temperatures, $\Theta\to 0$ and (\ref{utheta}) is solved 
for $k_\Theta\to 1/\sqrt{2}$, and so $q_{\Theta}^{+}\to\infty$.
To lowest order in $\Theta$, one has
$\cn\,\uT\approx\nc\,\uT\approx\dn\,\uT\approx 1$,
$\sn\,\uT\approx\uT$ and 
$E(\vT,k)\approx\vT\approx\uT$,
and so the term in curly brackets in (\ref{iqc}) vanishes
in order $\Theta$:
\begin{equation}
I[q_c]=\Theta\,U(q_t)+{\cal O}(\Theta^2).
\end{equation}
In $D(q_t,\Theta)$, Eq.\ (\ref{Dqt}), the third term in square
brackets behaves as $\uT^{-1}$, while all the others behave
as $\uT$ when $\Theta\to 0$. One has, therefore,
\begin{equation}
D(q_t,\Theta)=(2\pi g\Theta)^{-1/2}+{\cal O}(\Theta^{3/2}).
\end{equation}
It follows that
\begin{equation}
Z_2(\Theta)\stackrel{\Theta\to 0}{\sim}
\sqrt{\frac{1}{2\pi g\Theta}}\int_{-\infty}^\infty dq\, e^{-\Theta\,U(q)/g},
\label{z2theta0}
\end{equation}
or, equivalently,
\begin{equation}
Z_2(\beta)\stackrel{\beta\to 0}{\sim}
\sqrt{\frac{m}{2\pi\hbar^2\beta}}\int_{-\infty}^{\infty}dx\,e^{-\beta V(x)},
\end{equation}
with $V(x)$ and $U(q)$ defined in (\ref{vx}) and (\ref{ux}). This is, 
clearly, the ``classical'' limit for the partition function with a 
pre-factor that incorporates quantum fluctuations.


\subsubsection{Low temperatures}
\label{Lt}

At low temperatures, $\Theta\to\infty$ and (\ref{utheta}) is 
solved for $k_\Theta\to 1$. Using the asymptotic 
expansion [Ref.\ \cite{grads}, formula 8.113.1]
\begin{equation}
{\rm K}(k)=\ln{(4/k')}+\frac{1}{4}\,[\ln{(4/k')}-1]\,{k'}^2+\cdots,
\end{equation}
valid for $k'\equiv\sqrt{1-k^2}\to 0$, it follows
from (\ref{utheta}) that $k'_\Theta\approx 4\, e^{-\Theta/2}$ 
in leading order. Therefore,
\begin{equation}
q_{\Theta}^{+}=\sqrt{\frac{2{k'}_{\Theta}^2}{1-2{k'}_{\Theta}^2}}
\approx 4\sqrt{2}\,e^{-\Theta/2}.
\label{qT+}
\end{equation}

Since $|q_t|\le q_{\Theta}^{+}\ll 1$, one has 
$k\approx 1$, $\uT\approx\Theta/2$ and 
${\rm E}(\vT,k)\approx\sn(\uT,k)\approx\tanh(\Theta/2)\approx 1$ 
in the whole range of integration. One can also neglect
$\Theta\,U(q_t)$ and $q_t^2\uT$ in (\ref{iqc}), since both terms behave
as $\Theta\,e^{-\Theta}$. However, one must be more careful
with the combination $q_t^2\,\nc^2\uT\,(=q_0^2)$, since it grows
very rapidly with $q_t$ (in fact diverging when $|q_t|\to q_{\Theta}^{+}$),
and so cannot be treated as a formally small quantity.

Leaving this combination ``untouched'' in (\ref{iqc}), but making use of
the approximations listed above, we obtain
\begin{equation}
I[q_c]=\frac{4}{3}\left[\left(1+\frac{1}{2}\,q_t^2\,\nc^2\uT\right)^{3/2}
-1\right]+{\cal O}\left(\Theta\,e^{-\Theta}\right).
\label{ILT}
\end{equation}
The analysis of $D(q_t,\Theta)$ is much simpler. One can simply
take the limit $\Theta\to\infty$ of (\ref{Dg}), which was derived
under the assumption that $q_t\ll 1$.

Putting all pieces together we finally obtain 
\begin{equation}
Z_2(\Theta)\stackrel{\Theta\to\infty}{\sim}
\int_{-q_{\Theta}^{+}}^{q_{\Theta}^{+}}\frac{dq_t}{\sqrt{4\pi g}}\,
e^{-I[q_c]/g},
\label{Z2.1}
\end{equation}
with $q_{\Theta}^{+}$ and $I[q_c]$ given by (\ref{qT+}) and
(\ref{ILT}), respectively.

 
\subsection{Applications}
\label{applic}

We shall now apply the quadratic semiclassical approximation to obtain the 
ground-state energy and the curve for the specific heat as a function of 
temperature. These two applications will teach us about the usefulness of 
the approximation.
 
In order to compare (\ref{Z2.1}) with the expected low-temperature
limit of the partition function, 
$Z(\Theta)\sim e^{-\Theta\,\varepsilon_0(g)}$ 
(where $\varepsilon_0(g)\equiv E_0(g)/\hbar\omega$ is the dimensionless
ground state energy), it is convenient to rewrite it in a form in
which the $\Theta$-dependence can be analyzed more easily.
This can be done by changing the integration variable back to $q_0$.
Since $q_t\,\nc\,\uT=q_0$ and $q_{\Theta}^{+}$ is the value of $q_t$
corresponding to $q_0\to\infty$, one has
\begin{equation}
Z_2(\Theta)\stackrel{\Theta\to\infty}{\sim}
\int_{-\infty}^{\infty}\frac{dq_0}{\sqrt{4\pi g}}\left(
\frac{\partial q_t}{\partial q_0}\right)_{\Theta}\,
\exp\left\{-\frac{4}{3g}\left[\left(1+\frac{1}{2}\,q_0^2\right)^{3/2}
-1\right]\right\}.
\label{Z2.2}
\end{equation}
When $\Theta\gg 1$ it is possible to write an approximate 
expression for $q_t(q_0,\Theta)$, thus allowing to 
write the integrand in (\ref{Z2.2}) solely in
terms of $q_0$ and $\Theta$. The final result is
(see Appendix \ref{D} for details)
\begin{equation}
Z_2(\Theta)\stackrel{\Theta\to\infty}{\sim}\frac{2\,e^{-\Theta/2}}
{\sqrt{\pi g}}\int_{-\infty}^{\infty}dq_0\,\frac{
\exp\left\{-\frac{4}{3g}\left[\left(1+\frac{1}{2}\,q_0^2\right)^{3/2}
-1\right]\right\}}{\sqrt{1+\frac{1}{2}\,q_0^2}\left(1+\sqrt{1+\frac{1}{2}\,
q_0^2}\right)}.
\label{Z2.3}
\end{equation}
This gives $\varepsilon_0(g)=1/2$, indicating that the quadratic 
approximation is insufficient to yield corrections to 
the ground state energy of the harmonic oscillator. 
On the other hand, if one recalls that the partition function
can be written as
\begin{equation}
Z(\Theta)=\int_{-\infty}^{\infty}\rho(\Theta;q,q)\,dq,
\label{recall}
\end{equation}
where
\begin{equation}
\rho(\Theta;q,q)=\sum_{n}e^{-\Theta\varepsilon_n}\,|\psi_n(q)|^2
\stackrel{\Theta\to\infty}{\sim} e^{-\Theta\varepsilon_0}\,|\psi_0(q)|^2
\label{rho}
\end{equation}
is the diagonal element of the density matrix, one may take the
square root of the integrand in (\ref{Z2.3}) as an approximation
to the (unnormalized) wave function of the ground state. 
To test the accuracy of this approximation, we have
evaluated the expectation values of the energy for some values
of $g$ and compared them with high precision results found in the
literature. As Table \ref{T1} shows, the ground state energy
computed with this ``semiclassical'' wave function differs from
the exact one by less than $1\%$ even for $g$ as large as 2.

Another concrete problem that can be treated is the calculation of the
specific heat of the quantum anharmonic oscillator. 
It can be written in terms of $Z(\Theta)$ as
\begin{equation}
C=\Theta^2\left[\frac{1}{Z}\,\frac{\partial^2 Z}{\partial\Theta^2}
-\left(\frac{1}{Z}\,\frac{\partial Z}{\partial\Theta}\right)^2\right].
\label{SH}
\end{equation}
This expression was computed using MAPLE for a few values
of $\Theta$ and the coupling constant value $g=0.3$. 
The result is depicted in Fig.\ \ref{specheat},
which also exhibits the curve of specific heat of
the {\em classical} anharmonic oscillator (solid line).
As expected, the results agree when
the temperature is sufficiently high, but, in contrast to
the classical result, the semiclassical approximation
is qualitatively correct at low temperatures too,
dropping to zero as $T\to 0$.

This result, together with the estimate for the ground-state obtained 
previously, shows that the quadratic approximation works very well, being 
quite accurate at high temperatures, and still reliable at lower 
temperatures. In the next subsection, we will comment on why this is so.


\subsection{Beyond quadratic}

In this subsection, we shall compute the first correction $G_1$ 
to the quadratic approximation. For the sake of comparison, we shall use
both the quantum-mechanical (section \ref{qmpath}) and the field 
theory approaches (section \ref{ftpath}). 

In the quantum-mechanical approach, we want to compute
\begin{equation}
G_1(q_0,\Theta)=\int_0^\Theta d\theta\int_{-\infty}^\infty d\eta \,
G_c(0,0;\theta,\eta)\, \delta U[\theta,\eta]\, G_c(\theta,\eta;\Theta,0),
\end{equation}
with $\delta U$ given by (\ref{deltaU}), and $G_c$ constructed from 
(\ref{qatheta}) and (\ref{qtheta}), using (\ref{o12}) and (\ref{w12}).
Since $G_c$ is gaussian in $\eta$, and $\delta U$ is a polynomial
in $\eta$, the integral in $\eta$ can be readily performed,
yielding
\begin{equation}
G_1(q_0,\Theta)=\frac{3}{4}\,g^2\left[\frac{1}{\sqrt{4\pi g\,{\dot q}^2_c(0)
\,Q(0)}}\right]\int_0^\Theta d\theta\,\frac{{\dot q}^4_c(\theta)\,
[Q^2(\theta)-Q^2(0)]^2}{4Q^2(0)},
\label{deltagc}
\end{equation}
where the bracket in front can be identified with $G_c(0,0;\Theta,0)$. 

In the field-theoretic approach, we arrive at the same expression. 
In fact, the first correction corresponds to the $m=1$ term in 
(\ref{zpow}) which, using (\ref{deltaU}), yields
\begin{equation}
G_1(q_0,\Theta)=\int_0^\Theta d\theta\left[q_c(\theta)
\langle\eta^3(\theta)\rangle+\frac{1}{4}\,\langle\eta^4(\theta)
\rangle\right].
\end{equation}
Using (\ref{sumP}), we obtain
\begin{equation}
G_1(q_0,\Theta)=\frac{3}{4}\,g^2\,G_c(0,0;\Theta,0)\int_0^\Theta d\theta\, 
{\cal G}^2(\theta,\theta).
\label{delta1}
\end{equation}
($\langle\eta^3\rangle$ vanishes, and the factor 3 comes from the
three possible pairings of the four $\eta$'s in $\langle\eta^4\rangle$.)
Using (\ref{wronsk}) and (\ref{o12}) we reobtain (\ref{deltagc}).

Inserting (\ref{delta1}) in (\ref{z2.2}) and changing the integration
variable from $q_0$ to $q_t$ gives
\begin{equation}
Z(\Theta)=\int_{q^-_\Theta}^{q^+_\Theta} dq_t\, D(q_t,\Theta)\,
e^{-I[q_c]/g}\left[1-ga_1(q_t,\Theta)+\ldots\right],
\label{zcorr}
\end{equation}
where
\begin{equation}
a_1(q_t,\Theta)=\frac{3}{4}\int_0^{\Theta}d\theta\,
{\cal G}^2(\theta,\theta).
\label{a1}
\end{equation}
Because of the complicated form of ${\cal G}(\theta,\theta)$, 
it is not a simple task to compute $a_1(q_t,\Theta)$.
However, we can estimate the magnitude of this term
without much effort. Indeed, as shown in Appendix \ref{E},
${\cal G}(\theta,\theta)$ obeys the following inequality:
\begin{equation}
{\cal G}(\theta,\theta)\le\frac{\theta(\Theta-\theta)}{\Theta}
\qquad(0\le\theta\le\Theta).
\label{ineq}
\end{equation}
Therefore,
\begin{equation}
a_1(q_t,\Theta)\le\frac{\Theta^3}{40}.
\end{equation}
This shows that (in the case of the quartic anharmonic oscillator) 
the quadratic approximation to the partition
function, Eq.\ (\ref{z2.1}) or (\ref{z2.2}), can be used with
confidence whenever the condition $g\Theta^3/40\ll 1$ is
satisfied; this accounts for the numerical agreements obtained in the 
applications of the quadratic approximation.

A last comment is in order: the next term in the expansion for $Z(\Theta)$ 
has a piece with a factor $g$ and one with a factor $g^2$. The former 
comes from the product of $\langle\eta^6\rangle \sim g^3$ with the overall 
$g^{-2}$, whereas the latter involves $\langle\eta^8\rangle \sim g^4$. 
This is another indication that we are not dealing with a perturbative series.

\section{Conclusions}
\label{conclusions}

In order to understand the nature of the semiclassical series for the 
partition function, it is instructive to look at a diagrammatic expansion 
of $\rho(\Theta;q,q)$, which appears in (\ref{recall}). Comparing with 
(\ref{zbeta}), and defining 
$\rho_c(\theta_1,\eta_1;\theta_2,\eta_2)\equiv 
e^{-I[\theta_1,\theta_2;q_c]/g}\,G_c(\theta_1,\eta_1;\theta_2,\eta_2)$, 
we have the expansion in Fig.\ \ref{diag1},
where the dot represents the point $q$, the squares stand for the 
insertions of $[(-1)g^{-1}\delta U]$, the dashed lines for $\rho_c$, while 
the full line represents $\rho(\Theta,q,q)$. From 
(\ref{gc}) and (\ref{i2}), we may expand $\rho_c$ itself in terms of 
the density matrix elements for the harmonic oscillator (add and 
subtract $1$ to $U''[q_c]$ in (\ref{i2})), to obtain Fig.\ \ref{diag2},
where the circles represent insertions of $\{(-1)g^{-1}[U''[q_c]-1]\}$, 
and the dotted lines stand for the density matrix elements of the harmonic 
oscillator. Therefore, even the first term in our series already 
corresponds to an infinite sum of perturbative diagrams. Analogously, one 
could use the Feynman diagrams of the field-theoretic description to 
arrive at the same conclusion. This is a clear indication of the 
nonperturbative nature of our treatment.

The results of section \ref{statistical} can be generalized to 
higher-dimensional Quantum Statistical Mechanics, just as in 
Quantum Mechanics, where this was accomplished in 
\cite{morette1,mizrahi}. The generalization to 
potentials which allow for more than one classical solution, such as the 
double-well quartic anharmonic oscillator, requires a subtle matching of 
the series around each appropriate saddle-point (i.e., the minima). This is 
presently under investigation \cite{joras}.

An extension of our results to field theories is hampered by the fact 
that we do not know how to construct a semiclassical propagator in 
general. The technical simplifications which appear in one dimension cease 
to exist. However, our methods may still be of use in 
problems where classical solutions have a lot of symmetry (e.g., 
spherical symmetry) so that we can reduce them to effective 
one-dimensional problems. There are many such examples in Physics: 
instantons, monopoles, vortices and solitons are a few of the backgrounds 
that fall into that category. It is our intention to pursue this line 
of investigation.

Finally, we should remark that the field-theoretic treatment of 
subsection \ref{ftpath} can be used to compute any correlation function of 
interest. Therefore, a semiclassical series can be written down for any 
physical quantity once it is expressed in terms of the relevant correlations.


\acknowledgements

The authors acknowledge support from CNPq, FAPERJ and FUJB/UFRJ. 
RMC was supported in part by the National Science Foundation under 
Grant No.\ PHY94-07194. CAAC gladly acknowledges the support and 
hospitality of the ICTP, in Trieste, where substantial parts of this 
work were developed.


\appendix

\section{}
\label{A}

Here we show that $\eta_b(\theta)$, as defined in subsection \ref{qmpath},
is continuous at $\theta=\Theta/2$, and that it is also possible to
make $\dot\eta_b(\theta)$ continuous there by a suitable choice
of $Q(0)$.

To prove that $\eta_b(\theta)$ is continuous at $\Theta/2$,
we have to show that the limits $\lim_{\theta\uparrow\Theta/2}
\eta_b(\theta)$ and $\lim_{\theta\downarrow\Theta/2}\eta_b(\theta)$
exist and are equal. However, since $\eta_b(\theta)$ is even with respect 
to reflection around $\Theta/2$, it suffices to prove the existence
of one of them; the existence of the other and their equality will
then be automatically satisfied. Thus, let us consider the former
limit:
\begin{eqnarray}
\lim_{\theta\uparrow\Theta/2}\eta_b(\theta)&=&\lim_{\theta\uparrow\Theta/2}
\left[Q(0)\,\dot q_c(\theta)+\dot q_c(\theta)\int_0^{\theta}
\frac{d\theta'}{\dot q_c^2(\theta')}\right]
\nonumber \\
&=&0+\lim_{q\to q_t} v(q,q_t)\int_{q_0}^q \frac{dq'}{v^{3}(q',q_t)}.
\label{limleft}
\end{eqnarray}
To compute the above limit, we divide the interval of integration
in two subintervals, $[q_0,\tilde q]$ and $[\tilde q,q]$, with
$\tilde q$ close to $q_t$ but fixed. Then $v(q,q_t)$ times the
first integral vanishes when $q\to q_t$ and, by expanding $U(q)$
around $q_t$, we can approximate $v(q,q_t)$ times the
second integral by
\begin{equation}
\sqrt{2\,U'(q_t)(q-q_t)}\int_{\tilde q}^q\frac{dq'}
{[2\,U'(q_t)(q'-q_t)]^{3/2}}.
\end{equation}
Computing the integral and taking the limit, we finally obtain
\begin{equation}
\lim_{\theta\uparrow\Theta/2}\eta_b(\theta)=-\frac{1}{U'(q_t)}.
\label{qtilb}
\end{equation}

Let us now compute the limit
\begin{equation}
\lim_{\theta\uparrow\Theta/2}\dot\eta_b(\theta)=\lim_{\theta\uparrow\Theta/2}
\left[Q(0)\,\ddot q_c(\theta)+\ddot q_c(\theta)\int_0^\theta 
\frac{d\theta'}{{\dot q}^2_c(\theta')}+\frac{1}{{\dot q}_c(\theta)}\right].
\end{equation}
Using (\ref{euler}) and (\ref{motion}), and an integration by parts 
that cancels divergences, yields
\begin{equation}
\lim_{\theta\uparrow\Theta/2}\dot\eta_b(\theta)
=U'(q_t)\left[Q(0)+\frac{1}{U'(q_0)v(q_0,q_t)} 
-\int_{q_0}^{q_t}\frac{U''(q)\,dq}{[U'(q)]^2\,v(q,q_t)}\right].
\label{qtilbdot}
\end{equation}
(Although the integrand in (\ref{qtilbdot}) is singular as 
$q\to q_t$, the singularity is integrable.) 
Using (\ref{theta}) we may rewrite (\ref{qtilbdot}) as
\begin{equation}
\lim_{\theta\uparrow\Theta/2}\dot\eta_b(\theta)
=Q(0)\,U'(q_t)+\frac{1}{2}
\left(\frac{\partial\Theta}{\partial q_t}\right)_{q_0}.
\label{etadot}
\end{equation}
Since $\dot\eta_b(\theta)$ is odd with respect to reflection
around $\Theta/2$, $\lim_{\theta\downarrow\Theta/2}\dot\eta_b(\theta)
=-\lim_{\theta\uparrow\Theta/2}\dot\eta_b(\theta)$. However,
for $\dot\eta_b(\theta)$ to be continuous at $\Theta/2$, those
limits must be equal. This is possible only if 
$\lim_{\theta\uparrow\Theta/2}\dot\eta_b(\theta)=0$, or
\begin{equation}
Q(0)=-\frac{1}{U'(q_t)}\left(\frac{\partial\Theta}{\partial q_t}
\right)_{q_0}.
\label{id2}
\end{equation}


\section{}
\label{B}

In this appendix, we compute the pre-factor in (\ref{gc2}). To do so, 
we remark that (\ref{gc}) is the path-integral expression for the Euclidean 
time evolution operator $\hat\rho_c$ of the quantum-mechanical 
time-dependent problem defined by (note that the role of $\hbar$ is 
played by $g$)
\begin{equation}
-g\frac{\partial}{\partial\theta}\,\hat\rho_c(\theta,\theta')
=\hat H_c(\theta)\,\hat\rho_c(\theta,\theta'),
\label{quantum}
\end{equation}
with
\begin{equation}
\hat H_c(\theta)=-\frac{g^2}{2}\,\frac{\partial^2}{\partial q^2}
+\frac{1}{2}\,U''[q_c(\theta)]\,q^2.
\end{equation}
Indeed, we have
\begin{equation}
G_c(\theta_1,\eta_1;\theta_2,\eta_2)=\langle \eta_2 | 
\hat\rho_c(\theta_2,\theta_1)|\eta_1\rangle,
\end{equation}
so that $G_c(\theta_1,\eta_1;\theta_2,\eta_2)$ is a density matrix element.

Inserting (\ref{gc2}) into (\ref{quantum}), and using the fact that
$I_2$ satisfies the Hamilton-Jacobi equation
\begin{equation}
\frac{\partial I_2}{\partial\theta_2}+\frac{1}{2}\left(
\frac{\partial I_2}{\partial \eta_2}\right)^2-\frac{1}{2}\,
U''[q_c(\theta_2)]\,\eta_2^2=0,
\end{equation}
leads to
\begin{equation}
\left(\frac{\partial}{\partial\theta_2}+\frac{1}{2}\,
\frac{W_{12}}{\Omega_{12}}\right) G_c(\theta_1,0;\theta_2,0)=0,
\end{equation}
whose solution is (note that $W_{12}=\partial\Omega_{12}/\partial\theta_2$)
\begin{equation}
G_c(\theta_1,0;\theta_2,0)=C(\theta_1)\,\Omega_{12}^{-1/2}.
\label{gcap}
\end{equation}
$C(\theta_1)$ can be determined by demanding that we
recover the free-particle result as $\theta_1\to\theta_2$, 
\begin{equation}
G_c(\theta_1,0;\theta_2,0)\stackrel{\theta_1\to\theta_2}{\sim}
[2\pi g\,(\theta_2-\theta_1)]^{-1/2}.
\label{gclim}
\end{equation}
Expanding $\Omega_{12}$ around $\theta_1$, we obtain
\begin{equation}
\Omega_{12}=\Omega_{11}+\frac{\partial\Omega_{12}}{\partial\theta_2}
\Big|_{\theta_2=\theta_1}(\theta_2-\theta_1)+\ldots
=W_{11}\,(\theta_2-\theta_1)+\ldots
\end{equation}
Inserting this result in (\ref{gcap}) and comparing with (\ref{gclim})
we finally obtain
\begin{equation}
G_c(\theta_1,0;\theta_2,0)=\left[\frac{W_{11}}
{2\pi g\,\Omega_{12}}\right]^{1/2}.
\end{equation}


\section{}
\label{C}

Here we provide a simple argument to show why it is usually
simpler to write $q_0(q_t,\Theta)$ rather than $q_t(q_0,\Theta)$.
In fact, using $\zeta=q/q_t$ as integration variable in
Eq.\ (\ref{theta}), one has
\begin{equation}
\Theta=\int_1^{q_0/q_t}\frac{2|q_t|\,d\zeta}{\sqrt{2[U(\zeta q_t)-U(q_t)]}}
=F_{q_t}(q_0/q_t)-F_{q_t}(1),
\label{Fqt}
\end{equation}
where $F_{q_t}(\zeta)$ denotes the primitive of 
$2|q_t|/\sqrt{2[U(\zeta q_t)-U(q_t)]}$. One can then 
solve (\ref{Fqt}) for $q_0$, thus obtaining
\begin{equation}
q_0=q_t\,F_{q_t}^{-1}[\Theta+F_{q_t}(1)].
\end{equation}


\section{}
\label{D}

In this appendix, we derive an approximate expression for $q_t(q_0,\Theta)$
valid for $\Theta\gg 1$. We start by writing (\ref{theta}) with $U(q)$
given by (\ref{ux}):
\begin{equation}
\Theta=2\sqrt{2}\left(\int_{|q_t|}^a +\int_a^{|q_0|}\right)
\frac{dq}{\sqrt{(q^2-q_t^2)(q^2+q_t^2+2)}}\equiv I_1+I_2,
\end{equation}
where $a$ is a positive number such that $1\gg a\gg|q_t|$ (that both
inequalities can be simultaneously satisfied is guaranteed by the fact
that $|q_t|\le q_{\Theta}^{+}\approx 4\sqrt{2}\,e^{-\Theta/2}$
when $\Theta\gg 1$). Now, since both $a$ and $|q_t|$ are much smaller
than 1, one has $q^2+q_t^2+2\approx 2$ for $q$ in the interval $[|q_t|,a]$,
so that
\begin{equation}
I_1\approx 2\int_{|q_t|}^a\frac{dq}{\sqrt{q^2-q_t^2}}
=2\,\cosh^{-1}\left(\frac{a}{|q_t|}\right)
\approx 2\,\ln\left(\frac{2a}{|q_t|}\right),
\end{equation}
where the last (approximate) equality follows from $a/|q_t|$ being
much greater than 1.

In order to evaluate $I_2$, we use the fact that both $a$ and $|q_0|$
are much greater than $|q_t|$ to write
\begin{equation}
I_2\approx 2\sqrt{2}\int_a^{|q_0|}\frac{dq}{\sqrt{q^2(q^2+2)}} 
=2\,\ln\left(\frac{|q_0|}{a}\,\frac{1+\sqrt{1+\frac{1}{2}\,a^2}}
{1+\sqrt{1+\frac{1}{2}\,q_0^2}}\right)\approx
2\,\ln\left(\frac{2\,|q_0|/a}{1+\sqrt{1+\frac{1}{2}\,q_0^2}}\right).
\end{equation}
When adding the above results for $I_1$ and $I_2$, the dependence on
$a$ cancels out (as it should), and we are left with the following
result:
\begin{equation}
\Theta\approx 2\,\ln\left(\frac{4\,q_0/q_t}{1+\sqrt{1+\frac{1}{2}\,q_0^2}}
\right).
\end{equation}
This can be solved for $q_t$, yielding
\begin{equation}
q_t(q_0,\Theta)\approx\frac{4\,q_0\,e^{-\Theta/2}}
{1+\sqrt{1+\frac{1}{2}\,q_0^2}}.
\end{equation}
Using this expression to calculate the Jacobian in (\ref{Z2.2})
one finally obtains (\ref{Z2.3}).


\section{}
\label{E}

Let us define the function ${\cal G}_0(\theta,\theta')$ as
\begin{equation}
{\cal G}_0(\theta,\theta')=\cases{
\frac{\theta(\Theta-\theta')}{\Theta}, & $\theta\le\theta'$ \cr
\frac{\theta'(\Theta-\theta)}{\Theta}, & $\theta\ge\theta'$.}
\end{equation}
This function satisfies an equation similar to Eq.\ (\ref{green}):
\begin{equation}
-\frac{\partial^2}{\partial\theta^2}\,
{\cal G}_0(\theta,\theta')=\delta(\theta-\theta'),
\qquad{\cal G}_0(0,\theta')={\cal G}_0(\Theta,\theta')=0.
\label{green0}
\end{equation}
If we multiply (\ref{green}) by ${\cal G}_0(\theta,\theta')$ and
(\ref{green0}) by ${\cal G}(\theta,\theta')$, subtract one from
the other, and integrate the result from $\theta=0$ to $\theta=\Theta$,
we obtain
\begin{equation}
{\cal G}_0(\theta',\theta')-{\cal G}(\theta',\theta')
=\int_0^{\Theta}d\theta\,U''[q_c(\theta)]\,
{\cal G}_0(\theta,\theta')\,{\cal G}(\theta,\theta').
\label{G0-G}
\end{equation}
In the case of the quartic anharmonic oscillator, 
$U''[q_c(\theta)]=1+3q_c^2(\theta)>0$ and (consequently)
${\cal G}(\theta,\theta')\ge 0$ for $0\le\theta,\theta'\le\Theta$.
Since ${\cal G}_0(\theta,\theta')$ is also nonnegative in this
interval, Eq.\ (\ref{G0-G}) leads to the inequality (\ref{ineq}).



\begin{table}

\caption{Ground state energies for different values of $g$
($\hbar=m=\omega=1$).}
\label{T1}

\begin{tabular}{c c c c}
$g$ & $E_0$(semiclassical)\tablenote[1]{$\langle\phi_0|H|\phi_0\rangle/
\langle\phi_0|\phi_0\rangle$, where $\phi_0(q_0)$ is the square root
of the integrand in Eq.\ (\ref{Z2.3}).} 
& $E_0$(exact)\tablenote[2]{Values quoted from
Ref.~\cite{vinette}.} & error($\%$) \\
\hline
0.4 & 0.559258 & 0.559146 & 0.02 \\
1.2 & 0.639765 & 0.637992 & 0.28 \\
2.0 & 0.701429 & 0.696176 & 0.75 \\
4.0 & 0.823078 & 0.803771 & 2.40 \\
8.0 & 1.011928 & 0.951568 & 6.34 \\
\end{tabular}

\end{table}


\begin{figure}
\caption{$U(q)$.}
\label{potential}
\end{figure}

\begin{figure}
\caption{Graph of $f(k)$.}
\label{ktheta}
\end{figure}

\begin{figure}
\caption{Specific heat vs.\ temperature ($T=1/\Theta$) for the 
quantum (diamonds) and classical (solid line) anharmonic
oscillator. $g=0.3$.}
\label{specheat}
\end{figure}

\begin{figure}
\caption{Diagrammatic expansion for $\rho(\Theta;q,q)$.}
\label{diag1}
\end{figure}

\begin{figure}
\caption{Diagrammatic expansion for $\rho_c(\Theta;q_1,q_2 )$.}
\label{diag2}
\end{figure}


\end{document}